\documentclass[aps, prx, twocolumn, showpacs, superscriptaddress, floatfix, amsmath, amssymb]{revtex4-2}

\usepackage[utf8]{inputenc}
\usepackage{amstext}
\usepackage{amsfonts}
\usepackage{hyperref}
\usepackage{graphicx}
\usepackage{xcolor}
\usepackage[english]{babel}
\usepackage{microtype}
\usepackage{esint}
\usepackage{bm}

\usepackage{hyperref} 
\hypersetup{
	pdftoolbar=true,        		 
	pdfmenubar=true,     		 
	pdffitwindow=false,     	 
	pdfstartview={FitH}, 
	pdftitle={},  
	pdfauthor={},     		 
	pdfsubject={},   			 
	pdfcreator={},   	        	   	
	pdfproducer={}, 		
	pdfnewwindow=true,      	
	colorlinks=true,       		
	linkcolor=black,          	
	citecolor=blue,        		
	filecolor=magenta,    		
	urlcolor=blue          		
}

\usepackage{stmaryrd}
\newcommand{\CCQ}{Center for Computational Quantum Physics, Flatiron Institute, 162 5th Avenue, New York, NY 10010, USA}

\newcommand{\Grenoble}{Univ. Grenoble Alpes, CEA, Grenoble INP, IRIG, Pheliqs, F-38000 Grenoble, France}
\newcommand{\Neel}{Univ. Grenoble Alpes, CNRS, Institut N\'eel, 38000 Grenoble, France}

\renewcommand{\Re}{\text{Re }}
\renewcommand{\Im}{\text{Im }}

\begin{document}

\title{Cross-extrapolation reconstruction of low-rank functions and application to quantum many-body observables in the strong coupling regime}

\author{Matthieu Jeannin}
\email{matt.jeannin@gmail.com}
\affiliation{\Grenoble}

\author{Yuriel N\'u\~{n}ez-Fern\'andez }
\affiliation{\Neel}

\author{Thomas Kloss}
\affiliation{\Neel}

\author{Olivier Parcollet}
\affiliation{\CCQ}
\affiliation{Universit\'e Paris-Saclay, CNRS, CEA, Institut de physique th\'eorique, 91191,
	Gif-sur-Yvette, France}

\author{Xavier Waintal}
\email{xavier.waintal@cea.fr}
\affiliation{\Grenoble}

\date{\today}
	
\begin{abstract}
   
We present a general-purpose algorithm to extrapolate a low rank function of two variables 
from a small domain to a larger one.
It is based on the cross-interpolation formula.
We apply it to reconstruct physical quantities in some quantum many-body perturbative expansions in the real time Keldysh formalism, 
considered as a function of time $t$ and interaction $U$.
These functions are of remarkably low rank.
This property, combined with the 
convergence of the perturbative expansion in $U$ both at finite $t$ (for any $U$), and small $U$ (for any $t$),
is sufficient for our algorithm to reconstruct the physical quantity at long time, strong coupling regime.
Our method constitutes an alternative to standard resummation techniques in perturbative methods, such as diagrammatic Quantum Monte Carlo.
We benchmark it on the single impurity Anderson model and show that it is successful even in some regime
where standard conformal mapping resummation techniques fail.
\end{abstract}

\maketitle

\section{Introduction}

Many important problems in physics could formally be solved 
by extrapolating from a regime of parameters where 
physical quantities are known with a high degree of precision
to a more challenging regime. 
Common examples include extrapolating from weak to strong coupling, from large to small temperature or from short to long times.
Nevertheless, extrapolation is a notoriously difficult problem to control, 
unless it is based on some underlying rigid mathematical structure that strongly
constraints the extrapolation. Such a structure could take various form such as
an analytical structure in the complex plane \cite{Bertrand_1903_series, Bertrand_2021}, semi-positive definiteness \cite{Kemper_2023} or in the case of the present article, the low-rankness of some function. The approach that we will follow below is to control the extrapolation with - not just one but - two different variables.

Important extrapolation problems in computational quantum many-body physics
arise in the context of perturbative series expansions in power of the interaction.
The challenge is to resum them in strongly interacting regimes, 
i.e. beyond the {\it perturbative regime}
defined as the range of small interactions where the series actually converges.
Several methods have been used in the literature to do this, 
such as conformal mappings in the complex interaction plane \cite{Bertrand_1903_series, Conformalhistorical,Profumo2015,Rossi_2017,PhysRevLett.121.130405} 
Padé methods \cite{Pade-Historical, Bertrand_2021,baker1996pade}, or other mathematical approaches
\cite{Borel-Pade-Conformal,Van_Houcke_2019, Xseries,self-similar1,Self-similar2,PhysRevB.77.125101,Lindelof}.
Recently, several remarkable results have been obtained by combining these ideas 
with numerical methods to compute the series to relatively large orders,
both in equilibrium with diagrammatic quantum Monte Carlo (QMC) 
\cite{ProkofievSvistunov98, Vanhoucke2010,Van_Houcke_2012} 
and out of equilibrium \cite{Profumo2015,Nuniez_2022}. 
Let us mention for example an exact solution of a simple quantum dot model, 
in the long time out of equilibrium steady state or in the Kondo regime
\cite{Profumo2015, Bertrand_1903_series,Bertrand_1903_kernel, Nuniez_2022,macek2020qqmc,Bertrand_2021}, and an exact solution of the Hubbard model 
in a pseudo gap regime \cite{FerreroSimkovic2022,simkovic2022origin}.
Nevertheless, despite their numerous successes, these approaches have limitations and are difficult to automatize, 
for example in the context of quantum embedding solvers.
For instance, the singularities of the physical quantities 
in the complex plane of the interaction $U$ can limit our ability to use conformal mappings.

\begin{figure} 
	\centerline{ \includegraphics[scale=0.35]{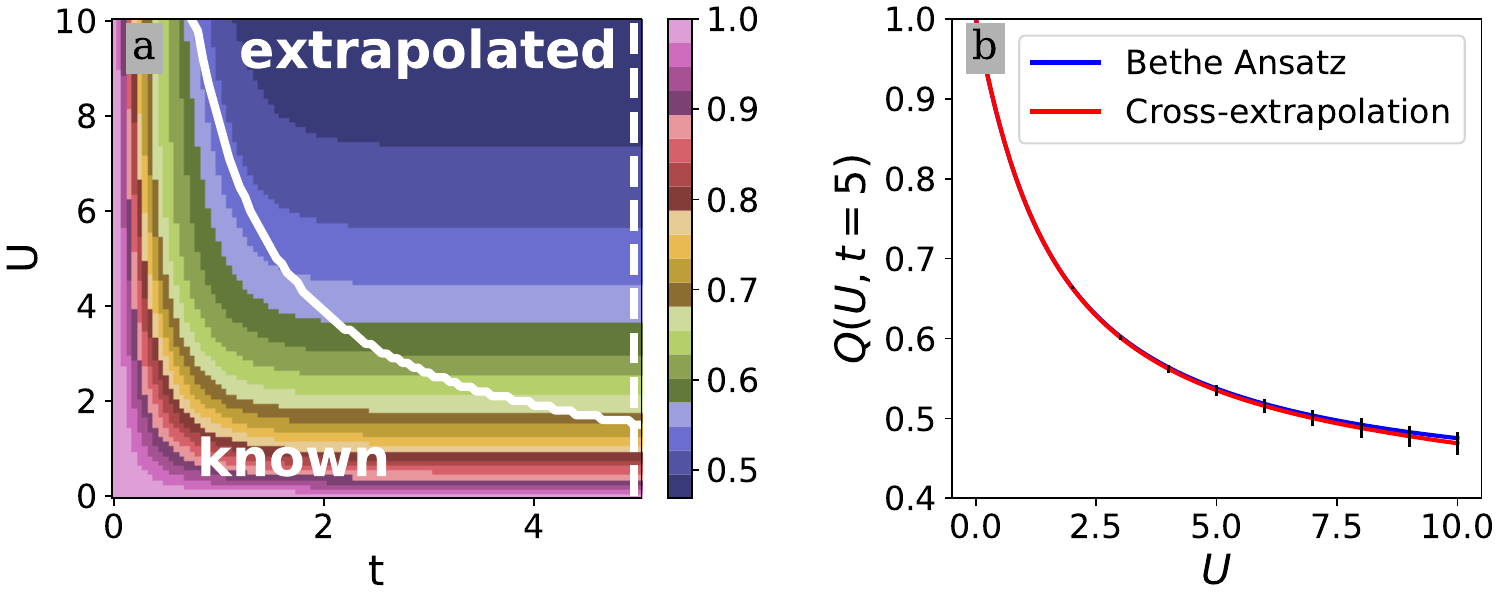} }
	\caption{\label{fig:QUt} Example of Cross-extrapolation. (a)
	   Charge $Q(U, t$) in the Anderson impurity model, at temperature $T=0$.
	   $U$ is the interaction, $t$ the time after switching of the interaction. 
	   The charge can be computed on the region below the white line (using an expansion in
	   power of $U$ to order 22); the region above the white line shows the double time-interaction cross-extrapolation. 
     (b) Comparison between the extrapolation at $t=5$ [along the dash line shown in (a)] 
       and the exact result (known from Bethe Ansatz). 
}
\end{figure}

In this paper, we follow a different route to reach the strong coupling regime. 
We consider a physical quantity $Q(U,t)$ that can be computed precisely for
large $t$ if $U$ is sufficiently small \emph{and} for relatively large $U$ if
$t$ is small enough. Here $U$ is the interaction strength and $t$ the time after an initial quench of the
interaction. Two kind of extrapolations could be attempted: as a function of $U$ (for
given $t$) or as a function of $t$ (for given $U$). Our ``cross-extrapolation'' algorithm performs
both extrapolations simultaneously, using the assumption that $Q(U,t)$ is low rank
[i.e. well approximated by a sum of a few terms of the form $f(t)g(U)$].
We shall show that this assumption is verified in concrete examples. 
Furthermore, this low-rank structure is sufficient to reconstruct the physical
quantities at large interactions and long times, based only on its value in the perturbative regime.
Let us mention that a different form of "low rankness" (using the quantics representation) has been exploited recently in a
- one variable - extrapolation scheme \cite{Lambert2023} to go beyond linear predictions. Both this scheme and ours
could possibly be combined (see \cite{Ritter2023} for a link between quantics and cross interpolation).

From a mathematical point of view, the problem is quite general and can be formulated as follows.
Given a function which is low-rank in some rectangular domain
(as defined precisely in Sect. \ref{sec:CE}), how to reconstruct it from its values in a subdomain ?
In Sect. \ref{sec:CE}, we discuss a simple {\it cross-extrapolation} algorithm to solve this problem,
based on the cross-interpolation (CI) decomposition of a matrix.
The algorithm is rank-revealing, i.e. it is able to reconstruct the function
but explicitly fails if the function is not of low rank. 

Our main result is illustrated in Fig.~\ref{fig:QUt}, which shows the charge of a
quantum dot in the Anderson impurity model in equilibrium, as a function of $U$
and $t$. The perturbative series is converged after 22 orders only below the white
line.
The low-rank extrapolation from this perturbative regime
to long times and large interactions is in excellent agreement at equilibrium with Bethe Ansatz benchmarks (Fig.~\ref{fig:QUt}b). Furthermore, the underlying property that permits the extrapolation (low-rankness)
is very different from the one used in conformal mapping method (positions of the poles and branch cuts in the complex $U$ plane \cite{Profumo2015, Bertrand_1903_series}). We explicitly exhibit an example where the former works
while the latter fail due to singularities located close to the positive real axis in the complex $U$ plane.

This paper is organized as follows.
In Sect. \ref{sec:CE}, we first present the cross-extrapolation algorithm and 
illustrate it on a simple toy function;
in section \ref{sec:ApplicationSIAM}, we apply this technique first to the charge of a quantum dot in equilibrium 
as a benchmark, and then to the current flowing through the dot upon applying a bias voltage; 
finally we conclude in section \ref{sec:conclusion}.


\section{Cross extrapolation}
\label{sec:CE}

In this section, we present our {\it cross-extrapolation} algorithm and 
test it on a simple toy function. 
The problem is the following: given a function $f(x,y)$ defined on domain ${\cal D} = [0,l_x]\times [0,l_y]$  but that is known only 
in a small subdomain ${\cal D}_{\rm sub} \subset {\cal D}$ (for instance for $xy < c$), we would like to extrapolate it to its entire domain ${\cal D}$. 

The extrapolation scheme requires the function to have a low-rank.
More precisely, $f$ is of $\epsilon$-rank $\chi$ 
(i.e. of rank $\chi$ up an error $\epsilon$)
if a set of $\chi$ one dimensional
functions $g_k$ and $h_k$ can be found such that
\begin{equation}
\label{eq:def_low_rank}
| f(x,y) - \sum_{k=1}^{\chi}g_{k}(x)h_{k}(y) |< \epsilon
\end{equation}
for all $x,y$. A function is of $\epsilon$-rank $\chi=1$ if it is almost factorizable.
This definition is the direct extension of the same concept for matrices.
If we use a discretization grid $(x_i, y_j)$ on the domain, with a step $\delta$
($|x_{i+1}-x_i|<\delta$) then the matrix $M_{ij} = f(x_i, y_j)$ is of $\epsilon$-rank $\chi$.
Conversely, if the matrix $M_{ij} = f(x_i, y_j)$ remains of $\epsilon$-rank $\chi$ in the continuous limit $\delta\rightarrow 0$ then the function $f$ is of $\epsilon$-rank $\chi$. We will explicitly 
checked in our applications that the result did not depend on the size of the grid used in the practical discretization.

\subsection{Matrix cross interpolation}
\label{section:CI}

Cross-extrapolation is based on the
{\it cross-interpolation} formula for matrices that we briefly recall \cite{matrixbook,GOREINOV19971,Kumar2017,HAMM20201088,Mahoney2009CURMD,matrix_completion}.
Given a matrix $M_{ij}$, we select a set of $\chi$ linearly independent lines of indices $i_\alpha$
($\alpha\in \{1\dots\chi\}$), and  $\chi$ linearly independent columns of indices $j_\beta$ ($\beta \in \{1\dots\chi\}$).
The $\chi\times\chi$ matrix formed by these two sets $P_{\alpha\beta} \equiv M_{i_\alpha,j_\beta}$ is called the
pivot matrix. The cross-interpolation $M^{\rm CI}$ of $M$ is defined as, 
\begin{equation}
\label{eq:defci}
M^{\rm CI}_{ij} \equiv \sum_{\alpha,\beta} M_{i,j_\beta}(P^{-1})_{\beta\alpha} M_{i_\alpha, j}
\end{equation}
Equation \eqref{eq:defci} has two main properties \cite{Nuniez_2022}:
{\it (i)} It is an interpolation, i.e. $M_{ij} = M^{\rm CI}_{ij}$ on the selected lines and columns
($i=i_\alpha$ or $j=j_\beta$);
{\it (ii)} if $M$ is exactly of rank $\chi$, the approximation is exact: $M=M^{\rm CI}$.
For a $\epsilon$-rank $\chi$ matrix, the cross-interpolation formula provides a practical way to obtain its low rank approximation. 
Crucially, the cross-interpolation is built purely from a few rows and columns
of the initial matrix. It \emph{does not} require the knowledge of the entire matrix. This is a sharp contrast with a
standard way of constructing a low rank approximation of a matrix, the singular value decomposition (SVD).
It can be shown \cite{maxvol1997,maxvol_2001,maxvol2010} that the optimal choice of the rows ($i_\alpha$)
and columns ($j_\beta$) corresponds to the maximization of the determinant $|P|$ of the pivot matrix.
In practice, the pivot selection is done using efficient heuristics that are much cheaper computationally
\cite{dolgov2020integralRn,Kumar2017,ACA2000,Bebendorf2011AdaptiveCA}. A standard choice is to 
build the approximation iteratively by looking for a pivot $(i_*,j_*)$ that currently holds a large error. 
This new pivot is added to the pivot list ($i_{\chi+1}=i_*$, $j_{\chi+1}=j_*$) to increase $\chi\rightarrow \chi+1$, 
see  Fig.~\ref{fig:annex_crossi} in Appendix \ref{sec:illustration} for an illustration.

Cross-interpolation extends naturally to functions of two variables $f(x,y)$ \cite{Kumar2017,ACA2000,matCIerror2010}.
Selecting $\chi$ pivots $(x_\alpha,y_\alpha)$ and defining the pivot matrix as  
$P_{\alpha\beta} \equiv f(x_\alpha,y_\beta)$,
we approximate $f(x,y)\approx  f_\chi^{\rm CI} (x,y)$ with its cross-interpolation approximation $ f^{\rm CI}_\chi (x,y)$,
\begin{equation}
\label{eq:ci_f}
{f}^{\rm CI}_\chi(x,y) \equiv \sum_{\alpha,\beta} f(x,y_\beta)(P^{-1})_{\beta\alpha} f(x_\alpha,y)
\end{equation}
This formula is often written using a slice notations (inspired by MATLAB notations): noting 
$\mathcal{X}$ the list of pivot rows $\{i_1\dots i_\chi\}$ and $\mathcal{Y}$ the list of pivot columns 
$\{j_1\dots j_\chi\}$, one writes
\begin{equation}
\label{eq:ci_f2}
{f}^{\rm CI}_\chi(x,y) \equiv  f(x,\mathcal{Y})f(\mathcal{X},\mathcal{Y})^{-1} f(\mathcal{X},y)
\end{equation}
The principle of the cross-extrapolation is very simple: we will extrapolate $f(x,y)$ using ${f}^{\rm CI}_\chi(x,y)$
where the pivots $(x_\alpha,y_\beta)$ are chosen inside the region ${\cal D}_{\rm sub}$ 
where $f(x,y)$ can be calculated. Property (ii) of the cross-interpolation guarantees that this extrapolation is exact if $f$ is exactly of rank $\chi$ and is
known with infinite precision.

\def\ftoy{f_{toy}} 

\subsection{Cross-extrapolation algorithm}
\label{section:CE}

\begin{figure} 
  
 \centerline{ \includegraphics[scale=0.52]{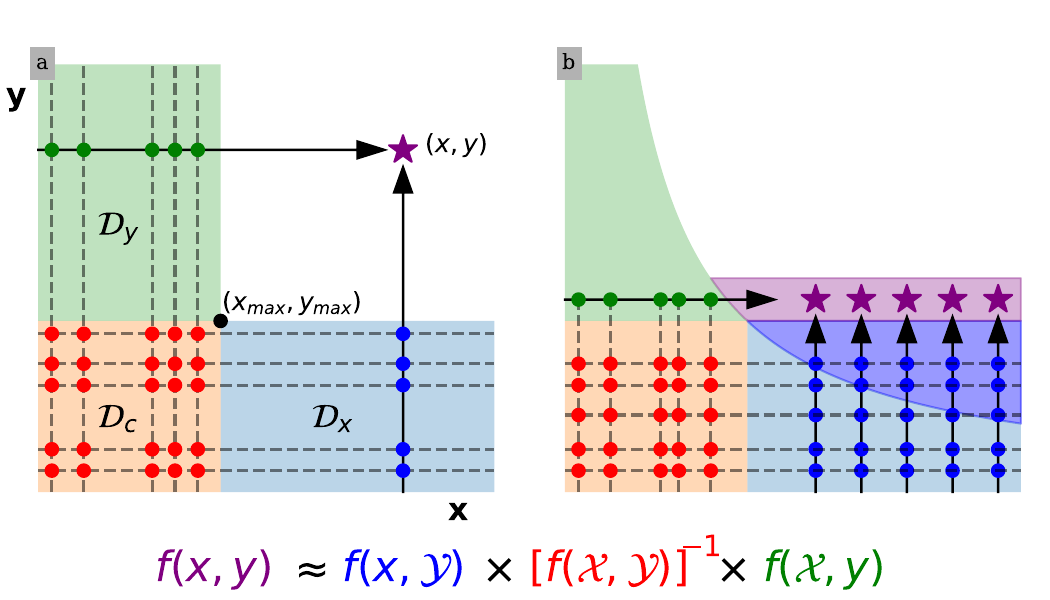} } 
\caption{\label{fig:ce}
Schematic of the cross-extrapolation algorithm.
{\it (a)} $L$-shape subdomain, cf text. The function is known in the colored regions, 
and the goal is to reconstruct it in the white region, e.g. at the $(x,y)$ represented 
by the purple star. Red circles correspond to the pivot matrix, blue and green ones to the column and the line of the target value.
{\it (b)} Hyperbolic subdomain. Each new line (purple)
     is extrapolated using information available (green, red and light blue) as
     well as the values newly extrapolated (deep blue) from the previous lines.
}
\end{figure}

\subsubsection{$L$-shaped subdomain}

We first consider the simple situation where the function is known on a
domain that has a $L-$shape (see Fig.~\ref{fig:ce}). The domain is made of three parts 
${\cal D}_{\rm sub} = {\cal D}_c \cup {\cal D}_x \cup {\cal D}_y $ 
where 
${\cal D}_c = [0,x_{max}]\times[O, y_{max}]$ (represented in orange in Fig.~\ref{fig:ce}), 
${\cal D}_x = [0,l_x]\times[O, y_{max}]$ (represented in blue) and  
${\cal D}_y= [0,x_{max}]\times[O, l_y]$ (represented in green).
The extrapolation is performed in two steps: first,
we perform a CI decomposition of $f$ inside ${\cal D}_c$ with $\chi$ pivots.
We obtain an approximation of the form \eqref{eq:ci_f} for $x,y \in {\cal D}_c$
where the pivots are in ${\cal D}_c$. Second, we use Eq. \eqref{eq:ci_f} to extrapolate 
 the function $f$ to the whole domain ${\cal D}$.
 Indeed, the calculation of ${f}^{\rm CI}_\chi(x,y)$ for any $x,y \in {\cal D}$ (e.g. the purple star point in 
Fig.~\ref{fig:ce}) uses only the values of $f$ on the $L-$shape domain ${\cal D}_{\rm sub}$. 

A central question in such an algorithm is the potential amplification of errors
in the extrapolation. In particular, we expect that if the subdomain is too small, 
the pivot matrix will become too singular (i.e. with very small singular value) to allow 
an accurate reconstruction. Indeed, even in the case where the extrapolation is mathematically exact 
(the function is exactly of rank $\chi$), the extrapolation is expected to fail in practice if the size of the subdomain becomes too small, if only because of the rounding error in e.g. double precision calculations
will ultimately get amplified by the fact that the pivot matrix will get highly ill conditioned. In real case calculations the function will be known with finite precision.
In order to illustrate this point, we benchmark the cross-extrapolation algorithm on a simple (toy) function, 
constructed to have a shape similar to the real case physical quantity that we will study later,
\begin{equation}
   \ftoy(x,y)=\left(\frac{x}{x+1}\right)^{4}(1+e^{-y^{2}})\left[1+y\cos(y)e^{-y\frac{x}{x+1}}\right]
\label{eq:f}
\end{equation}
where ${\cal D} = [0,l_x]\times [0,l_y]$ with $l_x=l_y=l=10$.
\begin{figure}
\centerline{ \includegraphics[scale=0.48]{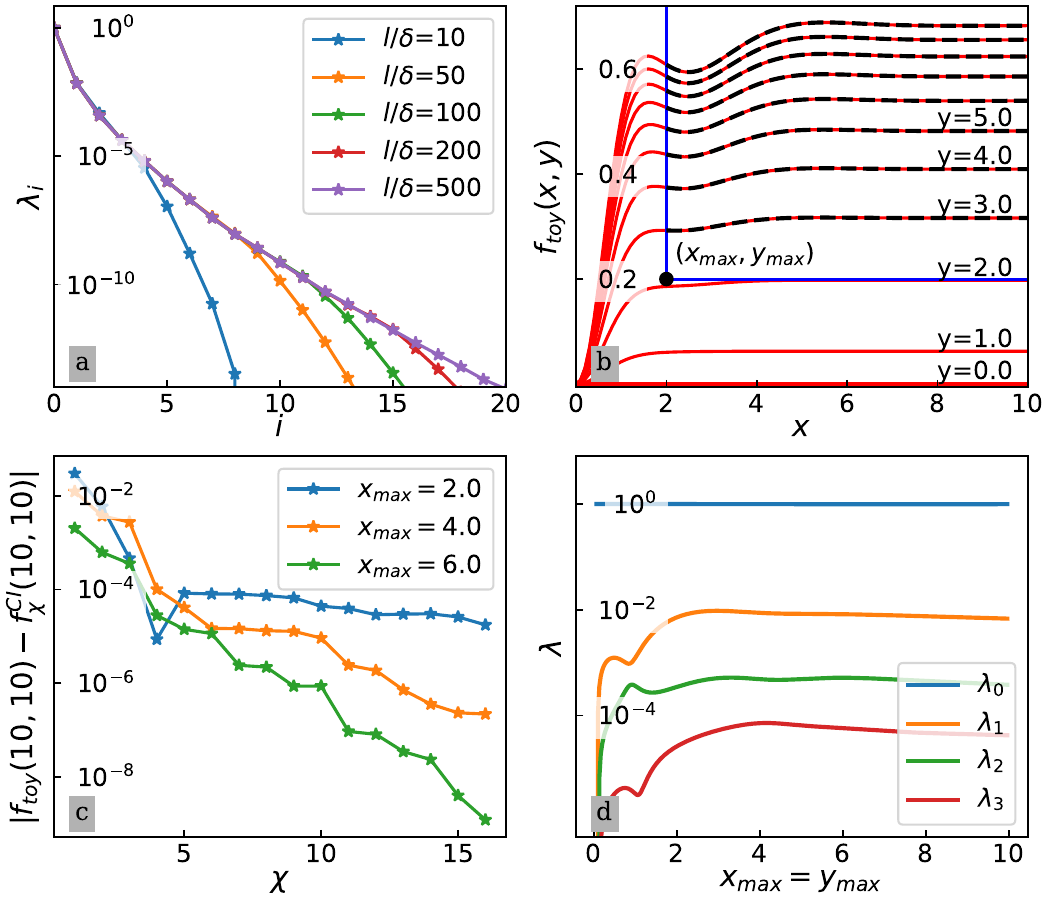} }
\caption{\label{fig:example_ce}
Benchmark of the cross-extrapolation algorithm on a $L$-shaped domain with $\ftoy$.
{\it (a)}
Singular values $\lambda_i$ (normalized by $\sum \lambda_i$) of $\ftoy$ vs $i$ for various numbers of grid points $l/\delta$ per dimension. In other panels, one sets $l/\delta=100$
{\it (b)}
Comparison of $\ftoy$ with its reconstructed value with cross-extrapolation 
for $x > x_{max}, y > y_{max}$, for  $x_{max}=y_{max}=2$, using  $\chi=5$ pivots;
the black dashed curve is the extrapolation, the red curve is $\ftoy$.
{\it (c)}
Error of the extrapolation at $x=y=10$ vs the rank $\chi$ for three values of $x_{max}$.
{\it (d)}
Four largest singular values of $\ftoy$ inside the subdomain ${\cal D}_c = [0,x_{max}]\times[0,y_{max}]$ versus $x_{max}$ for $x_{max}=y_{max}$.}
\end{figure}
The results are presented on Fig.~\ref{fig:example_ce}. 
First we check that the function $\ftoy$ is actually of low $\epsilon$-rank, 
by computing the SVD
of the matrix obtained by discretizing it on a
uniform grid with spacing $\delta$ ($l/\delta$ grid points per dimension). The (sorted) singular values $\lambda_i$ 
are shown on  Fig.~\ref{fig:example_ce}a, versus index $i$, for different grid discretization.
We observe that $\lambda_i$ decreases very quickly with $i$ (i.e. the function is of low rank 
to a very good approximation), and that they converge with the grid discretization $\delta\rightarrow 0$.

Second, we compare $\ftoy$ to the reconstructed function $f^{CI}_{\chi}$
with the cross-extrapolation from the $L$-shaped domain (with $x_{max}=y_{max}=2$) to the whole domain. 
Fig.~\ref{fig:example_ce}b shows a perfect match between the extrapolation (black dashed line) and the actual function (thin red) despite the fact that the extrapolation is non-trivial.

The corresponding extrapolation error $|\ftoy - f^{CI}_\chi|$ [for $(x,y)=(10,10)$] versus the rank $\chi$ is shown in Fig.~\ref{fig:example_ce}c. We observe a quick convergence of the error when $\chi$ is increased.
The convergence is faster when we use more information, i.e. for larger $x_{max}$.
A simple way to understand the role of the size of the subdomain (here given by $x_{max} = y_{max}$)
is to compute the SVD of the function (on a thin grid) on the ${\cal D}_c$ subdomain, as a function of $x_{max}$.
This is presented on Fig.~\ref{fig:example_ce}d.
For small $x_{max}$, the singular values are very small, increase with $x_{max}$ and then saturate.
If the subdomain is too small, the singular values of the function in it becomes very small, 
and the cross-extrapolation is expected to fail due to error amplification.
Qualitatively, Fig.~\ref{fig:example_ce}d provides a good \emph{a priori} indication of how well
cross-extrapolation will perform: in a favourable case, one can have access to values of $x_{max}$ large 
enough so that the singular values of the pivot matrix have already developed into a plateau; at the same time,
the faster the singular value $\lambda_i$ decays with $i$, the better.  See  Fig.~\ref{fig:annex_basic} in Appendix \ref{sec:illustration} for a step by step illustration of the cross-extrapolation from a $L$-shaped domain.

\subsubsection{General subdomain and error estimation}
\begin{figure}
\centerline{ \includegraphics[scale=0.45]{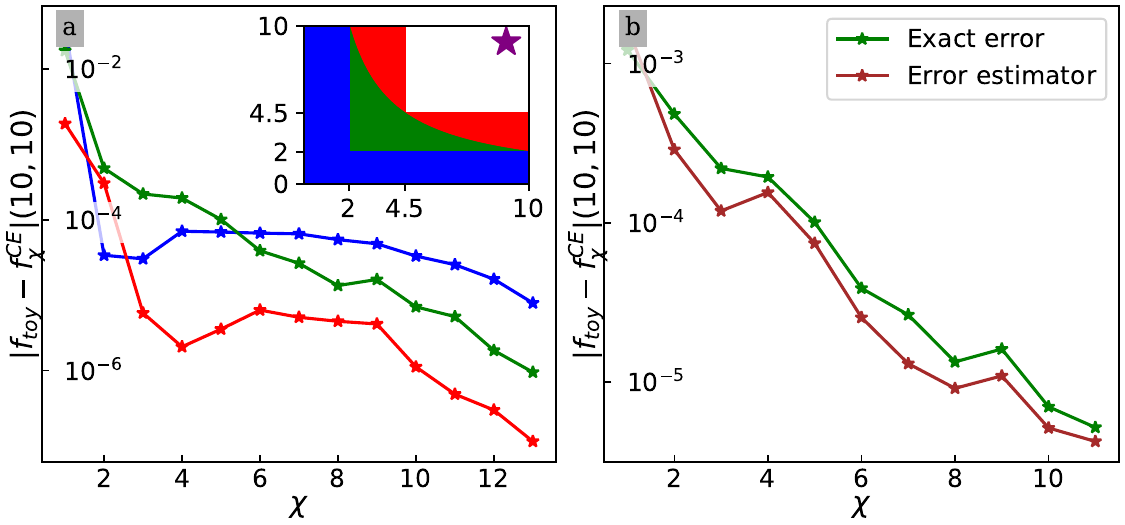} }
\caption{\label{fig:ce_opt_example}
{\it (a)}
Cross-extrapolation error at $(x=10, y=10)$ for three domains: an hyperbolic domain (green $\cup$  blue) and two $L$-shaped domains (blue,  red $\cup$ green $\cup$ blue).
The color of the curve correspond to the color of the domain.
{\it b)}
Error estimator $\epsilon_{\rm X}$ (red) and exact error (green) at $(x=10, y=10)$ as a function of $\chi$ for the green domain using $N_\chi = 3$.
}
\end{figure}

In practice, the subdomain ${\cal D}_{\rm sub}$ where the function can be
calculated can be more complex than a simple $L$-shape and we now extend the
cross extrapolation algorithm to handle a more general case.
We shall illustrate the general algorithm in the case where ${\cal D}_{\rm sub}$ is the area underneath an hyperbole
$xy < c$, where $c$ is some constant, which we will refer to as {\it hyperbolic subdomain},
see the schematic in 
Fig.~\ref{fig:ce}b.

Simply applying the $L$ shape algorithm here would be inefficient as it would not use all the available
information on the function. 
Our generalized algorithm works a follows (see Fig.
\ref{fig:ce}b for an illustration). We start from  ${\cal D}_x= [0,z]\times
[0,c/z]$ and use the $L$-shape algorithm to extrapolate the function to obtain
it just one line up (${\cal D}_x$ becomes ${\cal D}_x= [0,z]\times
[0,c/z+\delta]$). One continues the extrapolation up line by line, reusing the
extrapolated results of previous lines, until one has reached $y=z$. 
Crucially, the extrapolated points (in deep blue in Fig.
\ref{fig:ce}b) are reused only for the ${\cal D}_x$ domain of subsequent
extrapolations, but never for the ${\cal D}_c$ domain, as we observe that it would lead to instabilities. 
With this approach, the pivot region ${\cal D}_c$ is still rectangular, but evolves along the
computation to make use of the entire available information on the function. We
denote by $f^{\rm CE}_\chi$ the resulting Cross Extrapolation at rank $\chi$. 
Figure \ref{fig:annex_optimized} in Appendix \ref{sec:illustration} provides a step by
step illustration of the algorithm.

In Fig.~\ref{fig:ce_opt_example}, we present some benchmark using an hyperbolic domain for reconstructing $\ftoy$.
On the left panel, we show the error between the $\ftoy$ and $f^{CE}_\chi$, for 
two $L$-shape subdomains (blue and red) and the hyperbolic subdomain (green). 
As expected, we observe that the convergence is faster for larger subdomain at
large $\chi$: the errors are ordered in the opposite way compared to the inclusion relation
of the corresponding domains (blue $\subset$ green $\cup$ blue $\subset$ red $\cup$ green $\cup$ blue).
In our experiments, we have not observed any decrease of accuracy of the
generalized algorithm with respect to the $L$-shape algorithm (no amplification
of error due to the reusing of extrapolated values for subsequent
extrapolations). However, it is very important that at any stage the ${\cal
D}_c$ domain satisfies $xy<c$; indeed, the algorithm becomes unstable when one
uses pivots whose values have already been extrapolated (i.e. from the deep blue region of Fig.\ref{fig:ce}b).

Let us now discuss an estimator of the extrapolation error. 
In the simple toy model case where the function is known exactly in ${\cal D}_{\rm sub}$, the error is controlled by the rank $\chi$. We use $N_\chi$ consecutive values 
of $\chi$ to estimate it. In theory, $N_\chi=2$ is sufficient but it sometimes leads to fluctuations in the error due to accidental coincidence of the two estimates. In addition, $N_\chi=2$ supposes that the error is dominated by the next singular value, i.e. that they decrease rapidly. Hence, in practice we often use $N_\chi=3$ or $4$.
Anticipating on real case applications, a second source of error comes from the function being only known approximately inside ${\cal D}_{\rm sub}$ with an error $\eta$. Hence, we further generate $N_\eta$ extrapolations 
$f_{\chi,j}^{CE}(x,y)$ corresponding to $N_\eta$ different estimations of the function inside ${\cal D}_{\rm sub}$ ($N_\eta=1$ for the toy function). Our estimate of the error $\epsilon_{\rm X}$ of the extrapolation at point $(x,y)$ reads,
\begin{multline}
\epsilon_{\rm X} = \max_{i,i'\in\{0\dots N_\chi-1\} \atop j,j'\in\{1\dots N_\eta\}} |f_{\chi+i,j}^{CE}(x,y)-f_{\chi+i',j'}^{CE}(x,y)|
\end{multline}
Fig.~\ref{fig:ce_opt_example}b shows $\epsilon_{\rm X}$ versus $\chi$ for the toy function together with the
actual the exact error (deviation from the cross-extrapolation to the original function) which is known in the toy example. We find that the estimate is close to the exact error, and we will use this error estimator in the following.

The cross-extrapolation algorithm for two variables is {\it a priori} general.
Its success is linked to a property - low rankness - that we observe in the model below
and conjecture to be present in a large variety of similar situations.
Our numerical experiments suggest
that the method is robust: if we try it on a function which is {\it not} low
rank, we immediately observe a large increase of the error estimate.
Furthermore, the low rankness of the function can already be assessed
inside ${\cal D}_{\rm sub}$ using the analysis done in Fig.~\ref{fig:ce_opt_example}d.
Note however that, from a strictly mathematical point of view, 
the low-rank property of the function is not sufficient for the extrapolation to be controlled.
A simple failure example would be $f(x,y) = \theta(x-1) \theta(y-1)$ which can not be extrapolated 
from the sub-domain ${\cal D}_{\rm sub} = \{x<1\} \cup \{y<1\}$.

\section{Application to the extrapolation of time-dependant series}
\label{sec:ApplicationSIAM}

Let us now turn to the reconstruction of physical quantities from
their perturbative series at finite time $t$.
We study a well established model, the single quantum impurity Anderson model (SIAM) 
for a quantum dot. The SIAM is directly relevant experimentally, contains non-trivial Kondo physics,
is well understood at equilibrium (with exact benchmarks) while still under active study out-of-equilibrium
\cite{erpenbeck2023tensor,wauters2023simulations,Bertrand_2021,vanhoecke2023diagrammatic}.
\begin{figure} 
   \centerline{ \includegraphics[scale=0.35]{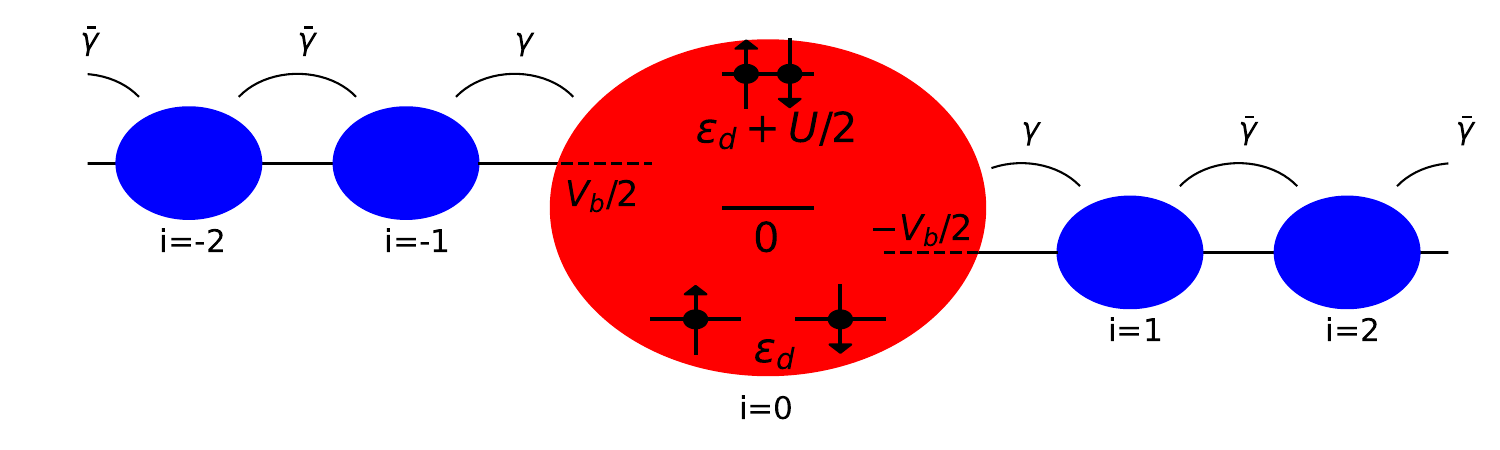} } 
   \caption{The Single Impurity Anderson Model :
   	a single-level quantum dot on site $i=0$ with on-site energy $\epsilon_d$ is subject to a finite Coulomb interaction $U$ and is hybridized with a tunnel coupling $\gamma$ with two semi-infinite leads ($i>0$ and $i<0$) that are biased with voltage $\pm V_b/2$. The hopping $\bar{\gamma}$ on  the electrodes is such that $\gamma\ll\bar{\gamma}$ } 
\end{figure}
The SIAM corresponds to a single interacting level (site $0$) weakly connected to two one dimensional leads
($i>0$ and $i<0$, see the schematic on Fig.~\ref{fig:siam}). Its Hamiltonian reads,
\begin{multline}
   \label{fig:siam}
   \hat{H}= \epsilon_d(\hat{n}_\uparrow+\hat{n}_\downarrow)+U\theta(t)
   (\hat{n}_\uparrow-\alpha)(
   \hat{n}_\downarrow-\alpha ) \\+ 
\sum_{i \in \mathbb{Z} \atop \sigma = \uparrow, \downarrow}
    \gamma_i c^\dagger_{i,\sigma
   }c_{i+1,\sigma } 
+
   \text{h.c.}
\end{multline}
where $c^\dagger_{i,\sigma}$ ($c_{i,\sigma}$) creates (destroys) an electron on site $i$ with spin $\sigma$,
$\epsilon_d$ is the on-site energy, $\hat{n}_{\sigma}\equiv \hat{c}^\dagger_{0,\sigma}\hat{c}_{0,\sigma}$ is the
density of electron in the impurity and $U$ the strength of interaction between
electrons which is switched on at $t=0$. 
$\gamma_i=\bar{\gamma}$ except for the coupling to the dot $\gamma_0=\gamma_{-1}=\gamma$.
The calculations are performed in the weak coupling limit $\gamma\ll\bar{\gamma}$ where the energy dependence of the
leads can be ignored (flat band approximation). The system can be put out-of-equilibrium in two ways:
using a symmetric voltage difference $\pm V_b/2$ across the leads and using a quench of the interaction. 
All times are written in unit of $\Gamma^{-1}$ with $\Gamma = 2\frac{\gamma^2}{\bar{\gamma}}$. 
The {\it shift parameter}
$\alpha$ does not affect the overall Hamiltonian as it can be absorbed in the on-site energy: 
$\hat H(\epsilon_d,\alpha)=\hat H(\epsilon_d-\alpha U,0)$ up to a constant energy shift. 
However, it allows one to perform several different expansions in power of $U$ to
reach the same final result (keeping $\epsilon_d-\alpha U=\text{const}$), but
with different radius of convergence at infinite time $t$ \cite{Profumo2015}.
\subsection{Charge of the quantum dot in equilibrium. Benchmark}
\begin{figure}
	\centerline{
		\includegraphics[scale=0.36]{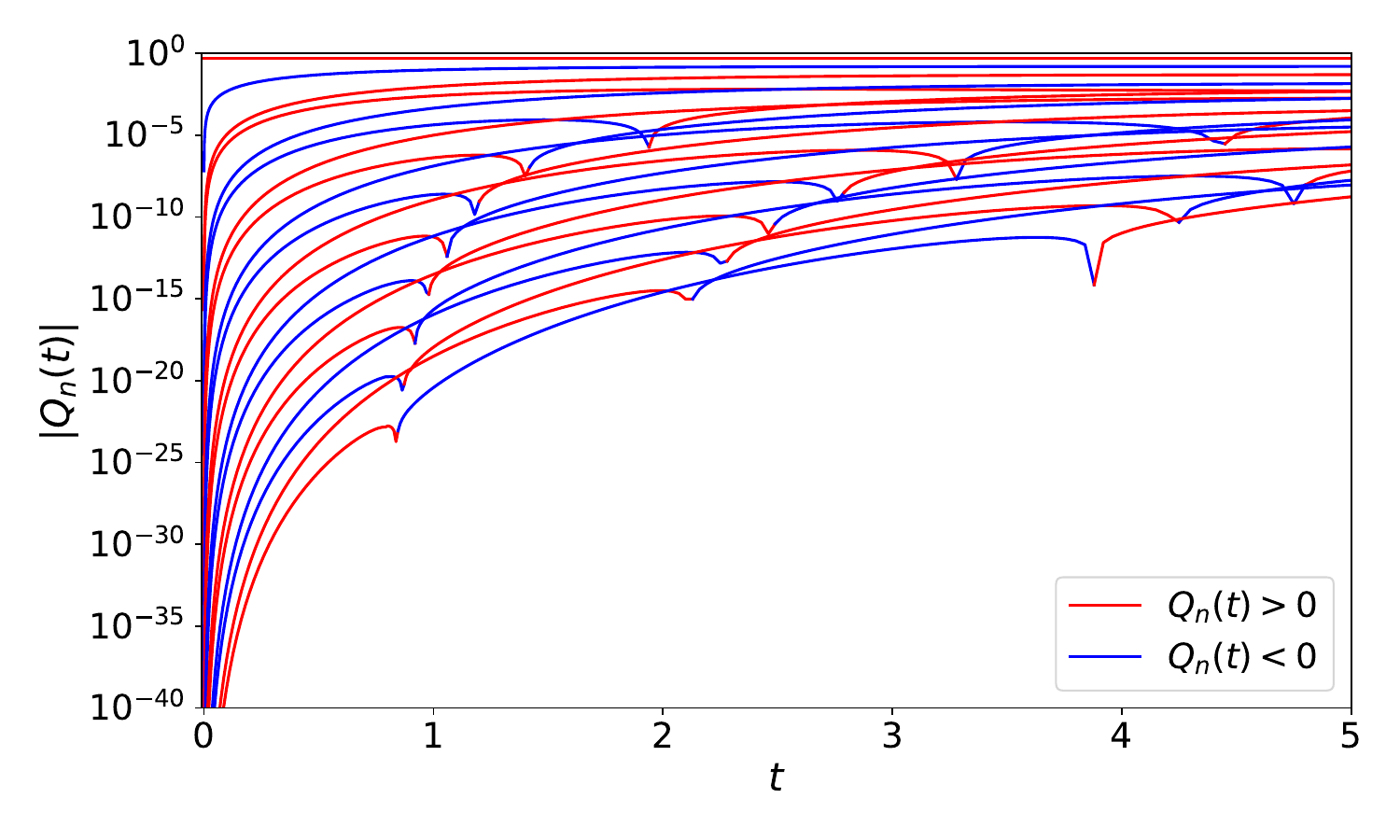}
	}
	\caption{Coefficients $Q_n(t)$ versus $t$ calculated with the TTD
	method for $\alpha=\epsilon_d=0$. From top to bottom the order $n$
     increases from $n=0$ to $n=19$. The color corresponds to positive (red)
  and negative (blue) values of the coefficient, the kinks to change of sign (the kinks get more pronounced upon adding more values of $t$ on the grid).
The relative error on the coefficients is of the order of $10^{-8}$ for $n=6$
and $10^{-5}$ for $n=19$, it is much smaller than the width of the lines. Note
the very large dispersion in magnitude of the coefficients. The coefficients
are shown up to $t=5$ but can be computed for arbitrary large $t$ at no
additional cost. Note that four different runs were needed in order to have
access to both large and small times with sufficient accuracy due to the
smallness of the coefficients at small time and large order (starting from $t=0$ with final times $t=0.01$, $t=0.1$, $t=1$ and $t=5$).
}
	\label{fig:raw_data}
\end{figure}

Let us first consider the charge on the quantum dot, at zero temperature $T=0$,
in equilibrium (no voltage bias). This quantity is a good benchmark for the 
technique, because we have Bethe Ansatz exact results for it at $t=\infty$ (see
Fig.~\ref{fig:QUt}) \cite{Wiegmann_Tsvelick} 
and it is an easy case of the conformal mapping
resummation technique \cite{Profumo2015, Bertrand_1903_series}.
In addition to this benchmark, we will also obtain the transient out-of-equilibrium
regime at finite time $t$ after the quench, a non-trivial calculation.
%
The charge is given by the expansion,
\begin{equation}
   \label{def:siam}
   Q(U,t)=\left<\hat{n}_\uparrow+\hat{n}_\downarrow\right>\approx \sum_{n=0}^{N} Q_n(t)U^n 
\end{equation}
where only a finite number $N$ of coefficients $Q_n(t)$ are known. The extrapolation problem consists in
extrapolating to the $N\rightarrow \infty$ limit. 
The calculation of the coefficient $Q_n(t)$ involves $n$-dimensional integrals
of a function consisting an exponentially large number of terms~\cite{Profumo2015, Bertrand_1903_kernel, Nuniez_2022}.
We use the tensor train diagrammatic (TTD) method within the real-time Keldysh formalism
to compute this expansion \cite{Nuniez_2022}.
Indeed, tensor network methods 
enable the computation of perturbative expansions of observables of this model to
unrivalled orders ($N\le 30$) and accuracies \cite{Nuniez_2022, erpenbeck2023tensor}.
Also, TTD provides the full time dependence in a single calculation \cite{Nuniez_2022}.
The coefficients $Q_n(t)$ calculated with TTD are shown in Fig.~\ref{fig:raw_data}. 
The frequent changes of sign indicates the presence of a sign problem that would make such a calculation at that precision
impossible with Monte-Carlo.
Also note the scale of the $y$-axis while the typical error bar is smaller than the width of the line.

\begin{figure}[th]
\centerline{ \includegraphics[scale=0.35]{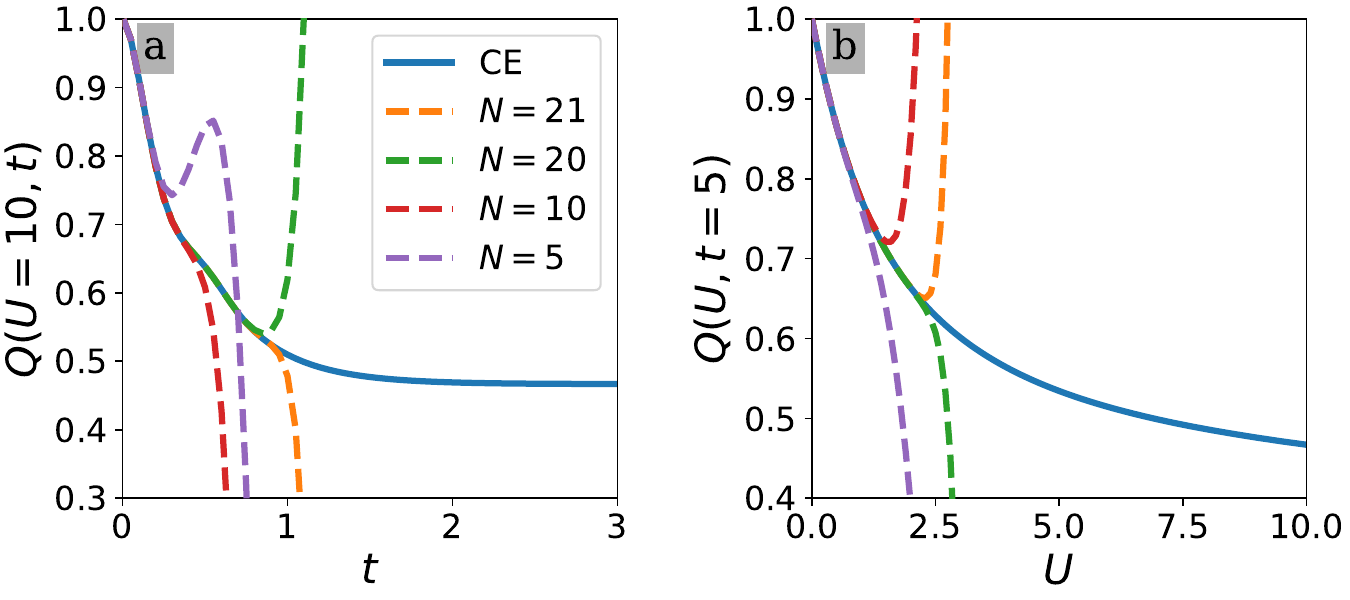} }
\caption{\label{fig:PropertySeriesQ}
Charge $Q(U= 10, t)$ vs $t$ (panel a), and  $Q(U, t = 5)$ vs $U$ (panel b)
calculated with bare summation (dashed lines, colors correspond to different values of $N$)  and the extrapolation using $N=21$ (plain blue line).
}
\end{figure}

At infinite time, the perturbative series has a finite radius of convergence $R$,
\begin{equation}
\lim_{t\rightarrow\infty} Q_n(t) \sim \frac{1}{R^n}
\end{equation}
and previous works have focused on the resummation of the series in this limit \cite{Bertrand_1903_series}.
However, at {\it any finite time}, the series has an infinite radius of convergence \cite{Bertrand_1903_series}, 
as the coefficients are bounded as 
\begin{equation}
Q_n(t) = O\left(\frac{t^n}{n!}\right).
\end{equation}  
For $U t< c$, the finite $N$ approximation is therefore converged 
for $N> c/ a$ up to an exponentially small error of order $(c/aN)^N$, 
where $a$ is a numerical coefficient of order unity.
These properties of the series are illustrated on Fig.~\ref{fig:PropertySeriesQ}, which shows, 
for the charge $Q(U, t)$, the exact value
(computed below), and the bare perturbative series summation 
for a finite number of order $N$.
At fixed large interaction $U$, the series is actually convergent, but the longer the time, 
the more orders are needed to achieve convergence, see Fig.~\ref{fig:PropertySeriesQ}a.
At infinite time, the series has finite radius of convergence, which manifests itself by a divergence of the finite $N$ sum at finite large time $t$, see Fig.~\ref{fig:PropertySeriesQ}b.

\begin{figure*}
\centerline{ \includegraphics[scale=0.45]{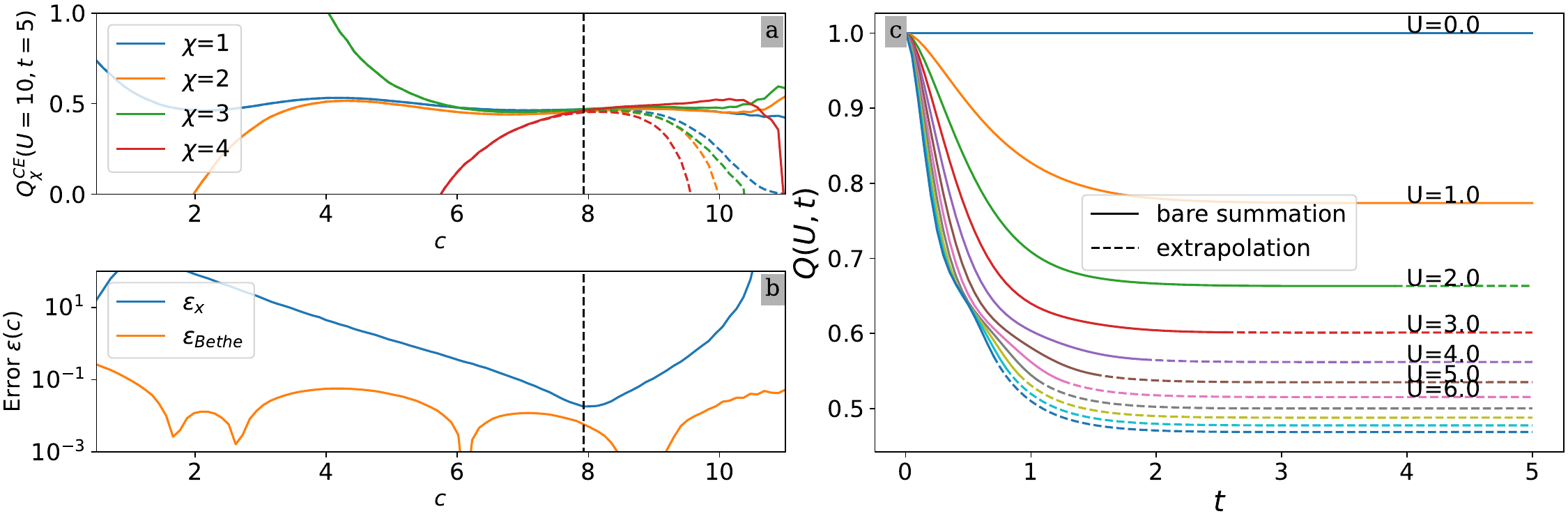} }
\caption{\label{fig:extrap_Qt}
{\it (a)} 
${Q^{CE}_\chi}(U=10,t=5)$ versus $c$, for $\chi = 1,2,3,4$ with $ \alpha=\epsilon_d=V_b=0$ and
$N=21$ (resp. $N =20$) coefficients of the perturbative series in plain (resp. dashed) line.
{\it (b)} Error estimator $\epsilon_x(c)$ (in blue) versus $c$ for the quantity plotted in panel {\it (a)}  and exact error $\epsilon_{Bethe}$ (orange). The exact error has been computed using the exact Bethe Ansatz series using the Euler conformal transformation method \cite{macek2020qqmc}.
The black dashed line marks the minimum of the error $c_{opt}$.
{\it c)} $Q(U,t)$ obtained by the cross-extrapolated algorithm.
Solid line sections of curves are in the perturbative regime $Ut<c_{opt}$ and are computed from the bare series, 
while dashed line sections are obtained by cross-extrapolation.
For all panels, we use a $100\times100$ grid and $c_{opt}=7.9$.
}
\end{figure*}

Our strategy is to compute $Q(U,t)$ in the hyperbolic domain ${\cal D}_{\rm sub}$ defined by
 \begin{equation}
 Ut<c
 \end{equation}
and use the cross-extrapolation algorithm to obtain $Q$ at larger times and interactions.
Since the computation of the coefficients in the Keldysh formalism scales as $2^n$, 
we have access to a limited number $N\le 30$ of them at best ($N\approx 30$ takes a few tens of hours and hundred computing cores, we use $N=20-25$ in practice for reasonable computational times) which sets the maximum value of $c$
that is actually reachable. The results of the extrapolation are shown in Fig.~\ref{fig:extrap_Qt}.
In order to check the stability of the algorithm, we vary the parameter $c$.
If $c$ is too small, as discussed earlier, ${\cal D}_{\rm sub}$ is too small, 
leading to an almost singular pivot matrix and unstable extrapolation. In practice, 
we obtain different results for different values of the rank $\chi$.
If $c$ is too large, the number of coefficients $N$ will be too small to converge the series
in ${\cal D}_{\rm sub}$ (hence varying $N$ yields different results).
The results are shown in Fig.~\ref{fig:extrap_Qt}a for various values of $\chi$ (different colors)
and two different values of $N$ (full versus dashed lines). As $c$ increases, we observe a convergence of the
extrapolated result. Using large value of $\chi$ requires large values of $c$ (here the $\chi=4$ curve is barely converged at $c=8$ while $\chi=2$ converges already at $c>5$) so the extrapolation will be most efficient for functions where small values $\chi=1$ or $2$ already provide a good extrapolation. For $c>8$, we do not have 
enough coefficients for convergence and the results become $N$ dependent.

In practice, we look for the optimal value of $c = c_{opt}$ for which the estimated error $\epsilon_{\rm X}(c)$ 
is minimal, see Fig.~\ref{fig:extrap_Qt}b ($c_{opt}\approx 8$ in this example). 
Here we have used $N_\chi=3$ and $N_\eta=2$ (two values $N=20$ and $21$)
for the calculation of $\epsilon_{\rm X}(c)$. Notice that our error estimate (blue) is conservative versus the true
error (orange) that we can obtain for this benchmark using Bethe Ansatz. The final extrapolation versus time $t$
is presented for different $U$ in Fig.~\ref{fig:extrap_Qt}c  (see also Fig.~\ref{fig:QUt}a for a 2D colorplot
and Fig.~\ref{fig:QUt}b for the large time limit versus $U$, in quantitative
agreement with the Bethe Ansatz benchmark). While the asymptotic result was
known already from previous work, the full curve Fig.~\ref{fig:extrap_Qt}c is a
non-trivial out-of-equilibrium calculation of an interaction quench. 
Notice the kink that seems to develop around $t\approx 0.8$ at large $U$ (whose physical
interpretation we leave to future work).

\begin{figure}[th]
	\centerline{ \includegraphics[scale=0.35]{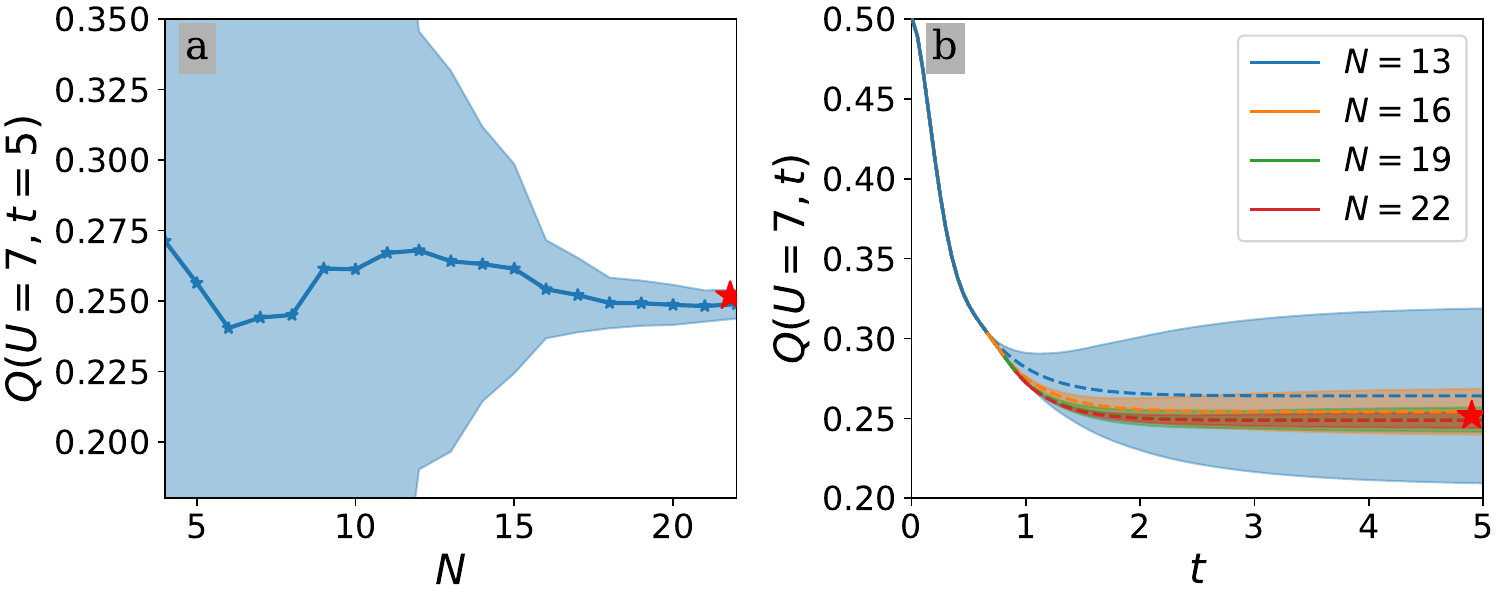} }
	\caption{\label{fig:DiffN} {
\it (a)} 
Charge $Q^{CE}$ vs $N$, for $\alpha=\epsilon_d=V_b=0$ at $U=7$ and $t=5$.
Error bars from CE are given by the shaded area. The red star is obtained using conformal maps. 
{\it (b)} 
Charge $Q^{CE}$ vs $t$, for $U=7$ and different values of $N$. 
Plain (resp. dotted) lines correspond to bare summation (resp. extrapolation).}
\end{figure}

An important property of the cross-extrapolation approach (as opposed to conformal transform as we shall discuss below) is that it
can be systematically improved by computing more orders.
This is checked in Fig.~\ref{fig:DiffN} where we study the influence of $N$ (the number of orders) by plotting the charge obtained by cross-extrapolation $Q^{CE}$ for different values of $N$ and time $t$. 
We observe that the error bar quickly decreases with $N$ for a given time, see Fig.~\ref{fig:DiffN}a.
As a function of $t$, we see that the error is small at short time, larger at long time, but also significantly decreases when $N$ increases, see Fig.~\ref{fig:DiffN}b.


\subsection{Current through the quantum dot}

\begin{figure*}
\centerline{	
\includegraphics[scale=0.45]{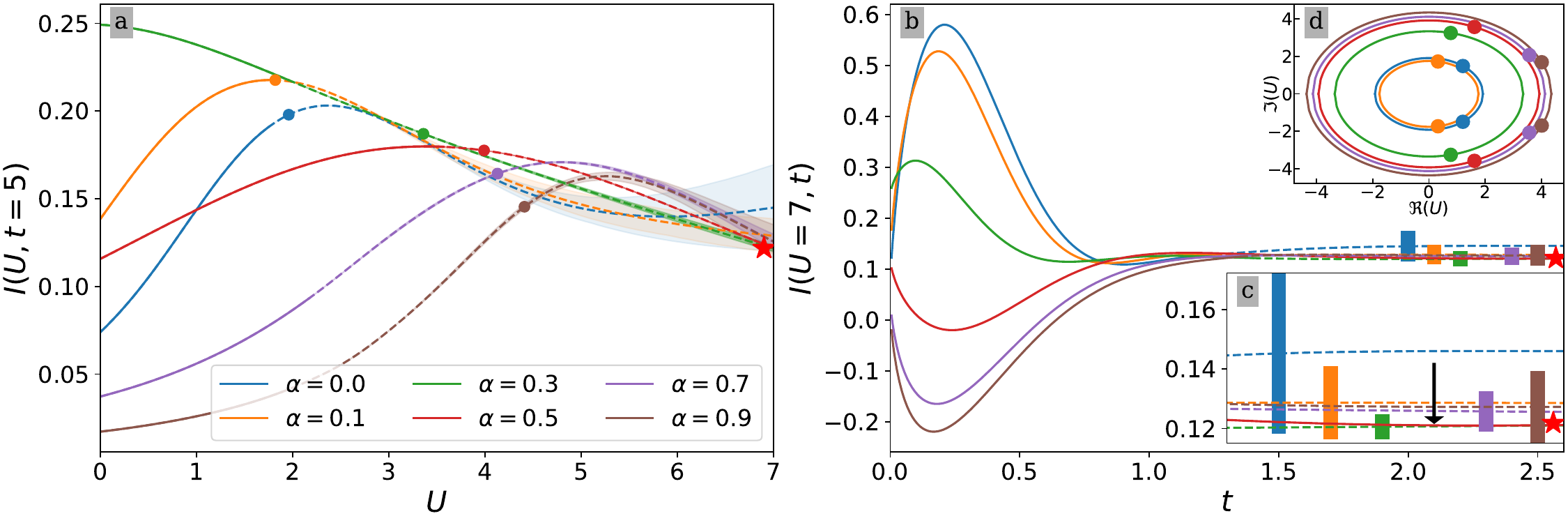} }
\caption{\label{fig:extrap_It}
{\it (a)} 
Current vs $U$ for various values of $\alpha$ and $\epsilon_d$ so that the on-site energy at $U=7$ verifies $\epsilon_d -7\alpha =2$ with bias voltage $V_b = 2$ and using $N=21$ coefficients. 
The different curves should and do converge to a single point at $U=7$ (the red star). 
Full lines correspond to  the simple plain summation of the $N$ coefficient; they turn into dashed lines when the extrapolation starts. 
The circles indicate the position of the radius of convergence of the corresponding series at infinite time (above this point the bare series diverges). 
The shaded areas indicate the error bars.  
{\it (b)} 
Same data as a function of time $t$ for $U=7$ and various $\alpha$ values. 
The color rectangles show the error bars for the curve of the corresponding color. The error for the red curve $\alpha=0.5$
is small and not visible at this scale 
{\it (c)} zoom of (b) in the extrapolation region. Note that $\alpha=0.5$ is the most precise prediction (the error bar is of the order of the line width at the position indicated by the arrow) and $\alpha=0$ the worst. 
{\it (d)} Position of the poles of the asymptotic series in the complex plane $(\Re(U),\Im(U))$ for the same series in the large $t$ limit.}
\end{figure*}

We now turn to the calculation of the current flowing through the quantum dot to the right electrode. 
The interaction quench takes the system from a  non-equilibrium non-interacting steady state to 
an non-equilibrium interacting steady state.
We want to compute both the steady state in the presence of voltage at long time, 
and its transient regime after the interaction quench.
Since we have no benchmark here, we perform several calculations varying the shift parameter $\alpha$,
using different pairs $(\epsilon_d,\alpha)$
such that $\epsilon_d -7\alpha =2$,
which must produce the same result at $ U=7$.
The calculation of the current $I(V_b=2,\epsilon_d=2,\alpha=0,U=7)$ is presented in 
Fig.~\ref{fig:extrap_It}a.
All calculations of $I$ indeed coincide (the red star) within their respective error bars. 
Some extrapolations have relatively large error bars
(the worst is $\alpha=0$ which becomes rather inaccurate for $U>4$), others are
more precise (the best is $\alpha=0.5$) and could be pushed to much larger
values of $U$ (above $U=10$ in this case).
 
Fig.~\ref{fig:extrap_It}b show the same data versus time for the different
quenches and $U=7$. All the different quenches converge to the same value, as
they should, within the calculated error bars (see Fig.~\ref{fig:extrap_It}c for
a zoom of the asymptotic regime, the error bar for $\alpha=0.5$ is too small to
be visible). The results of Fig.~\ref{fig:extrap_It} show that we have now
access to out-of-equilibrium observables, away from the perturbative regime,
and for a wide range of parameters (before 
\cite{Nuniez_2022}, only a narrow region in $\epsilon_d$ could be computed). We
defer to a subsequent publication a full study of the physics made available by
this combination of TTD and the cross-extrapolation.

Let us now compare the cross-extrapolation with the technique used in previous works, conformal transformation \cite{Bertrand_1903_series}.
Conformal transformation uses the series at $t=\infty$, 
\begin{equation}
   \label{eq:assympt}
   Q(U,t=\infty)= \sum_{n=0}^{\infty} Q_n(t=\infty)U^n 
\end{equation}
which we have also calculated for the same parameters (in practice, the calculations are performed at a finite time that satisfies
$t\gg n$, the computation time does not depend on this choice).
The series has a finite radius
of convergence $R$, hence the summation in 
Eq.\eqref{eq:assympt} diverges for $U>R$ as one increases the number of
coefficients $N$ kept in the calculation. 
The strategy of the conformal transform is to locate the poles in the complex
$U$ plane  $(\Re(U),\Im(U))$ that are responsible for the finite radius of
convergence, then design a conformal transformation that send these poles away
while bringing the value of $U$ for which we want to
calculate the observable ($U_{phys}$) closer to the origin \cite{Bertrand_1903_series}. 
In our experience, the
conformal transform yields spectacular results when the poles are situated on
the left of the origin $\Re(U)<0$ when one wants to calculate the observable
for $\Re(U)>0$: in that case, there is a clear separation between the poles and
$U_{phys}$ and $N=10-12$ coefficients with $1\%$ accuracy can be
sufficient to calculate the observable up to large $U$, even $U=\infty$
\cite{Profumo2015}. 
However, when the poles get closer to the points of interest (the poles lie on
the right of the origin $\Re(U)>0$), the conformal transform fails (but not
silently). In such case, computing more orders (up to
$N\approx 25$ as in the present work) or even radically improving the accuracy of the
calculation (up to $10^{-8}$ relative precision in the present work compared to
$10^{-2}-10^{-3}$ with Monte-Carlo sampling) does not really change the
outcome.

Remarkably, the cross-extrapolation can succeed when the conformal technique fail.
The computation of the current $I$ illustrates this.
Fig.~\ref{fig:extrap_It}d shows the positions of the poles of $Q(U, t=\infty)$ in the
$(\Re(U),\Im(U))$ plane, and the different radius of convergences for various $\alpha$. 
The poles lie in the $\Re(U)>0$ part of the plane. As a result, we cannot
perform the resummation using conformal transformations in this case, 
while the cross-extrapolation works perfectly.
%
Furthermore, cross-extrapolation systematically takes advantage of
more orders (increase the time up to which the observable can be computed) and
more accuracy (increases the range where the observable can be extrapolated).

\section{Conclusion} 
\label{sec:conclusion}

We have shown that some physical quantities on the non-equilibrium Anderson model, 
when considered as a function of $U$ and $t$ are of low-rank.
Furthermore, perturbative techniques give us access the short time regime (for a large range of $U$)
and the small interaction regime (for any time $t$). 
Due to the low-rank property, we have shown that the knowledge of the function 
in these two regimes is sufficient to extrapolate it reliably at long time and strong coupling.

In order to perform this calculation, we have presented a general-purpose {\it cross-extrapolation}
algorithm to extrapolate a low rank function of two variables from a small domain to a larger one, 
i.e. perform \emph{two-variables} extrapolations.
In our experience, this algorithm is rank-revealing i.e. it does not fail silently: 
when the function is not low rank or the computed domain is too small, the error bar simply increases until the extrapolation is no longer useful.
We expect this algorithm to be of general use, beyond the special case discussed in this paper, in condensed matter physics or beyond.
It could also potentially be extended to more than two variables using the tensor cross interpolation \cite{Nuniez_2022}. 

We have shown that the cross-extrapolation is able to reconstruct some physical
quantities at strong coupling in a case where the previously established
approach based on conformal maps~\cite{Profumo2015} fails.  Furthermore, in contrast with the conformal maps,
it can be systematically improved by calculating more coefficients (although the
computational cost scales as $2^N$) and it can be largely automatized,
which is crucial for potential future application to self-consistent methods such as non-equilibrium dynamical
mean field theory \cite{Georges1996}.

Finally, we conjecture that the crucial low-rank properties of physical quantities in the  $(U,t)$ plane
is a general properties of similar models. In particular, in the context of equilibrium diagrammatic QMC, 
it will be very interesting to examine the rank of physical properties in terms of interaction $U$ and inverse temperature $\beta$, 
as we expect $\beta$ to play a similar role of infra-red regulator as the time $t$ in the out-of-equilibrium context.
In practice, we could even use two different but complementary techniques, one to compute the small $U$ regime and another one to compute the small $\beta$ regime.

\section{Acknowledgements} 
The Flatiron Institute is a division of the Simons Foundation.
XW acknowledges funding from the Plan France 2030 ANR-22-PETQ-0007 “EPIQ”, from the French ANR DADDI and from the
AI program of the French MESRI.
\bibliography{refs}


\clearpage

\appendix

\section{Illustration of cross-interpolation and cross-extrapolation}
\label{sec:illustration}
\begin{figure}
	\centerline{	\includegraphics[scale=0.47]{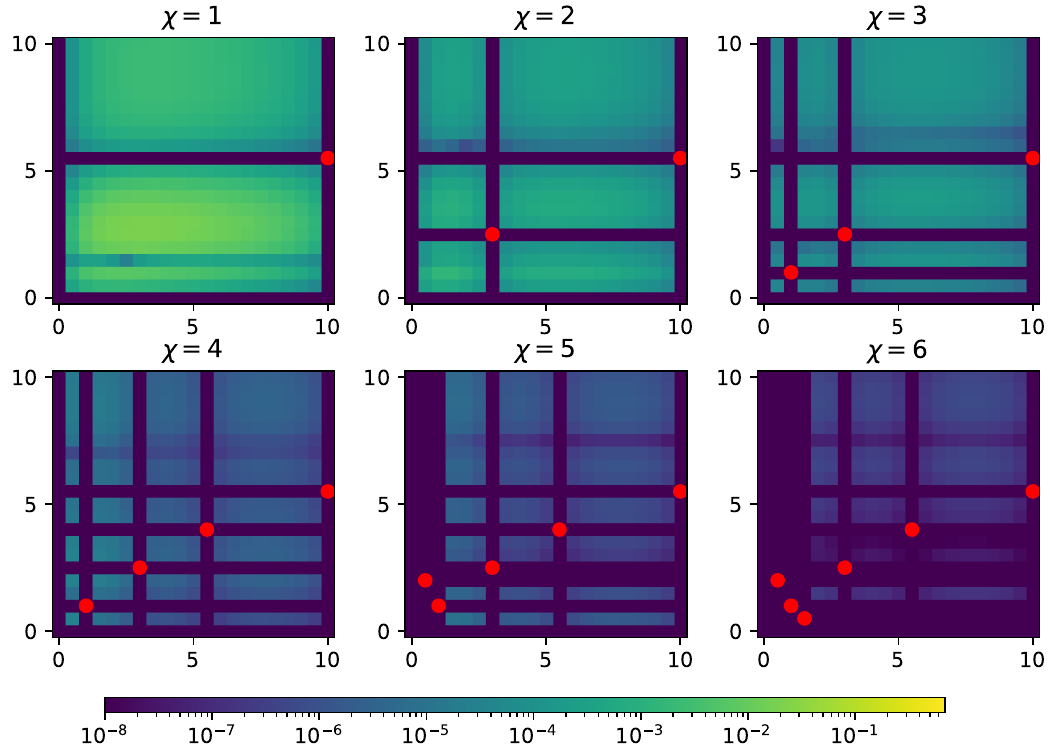}}
	\caption{Error versus $x$ and $y$ at different stages of the Cross-Interpolation. The illustration is
	done on the function $\ftoy$ defined in Eq. \eqref{eq:f} on a uniform $20\times20$ grid for $(x,y)\in [0,10]^2$.
	The red dots indicate the pivots. }
	\label{fig:annex_crossi}
\end{figure}

\begin{figure}		
	\centerline{
	\includegraphics[scale=0.47]{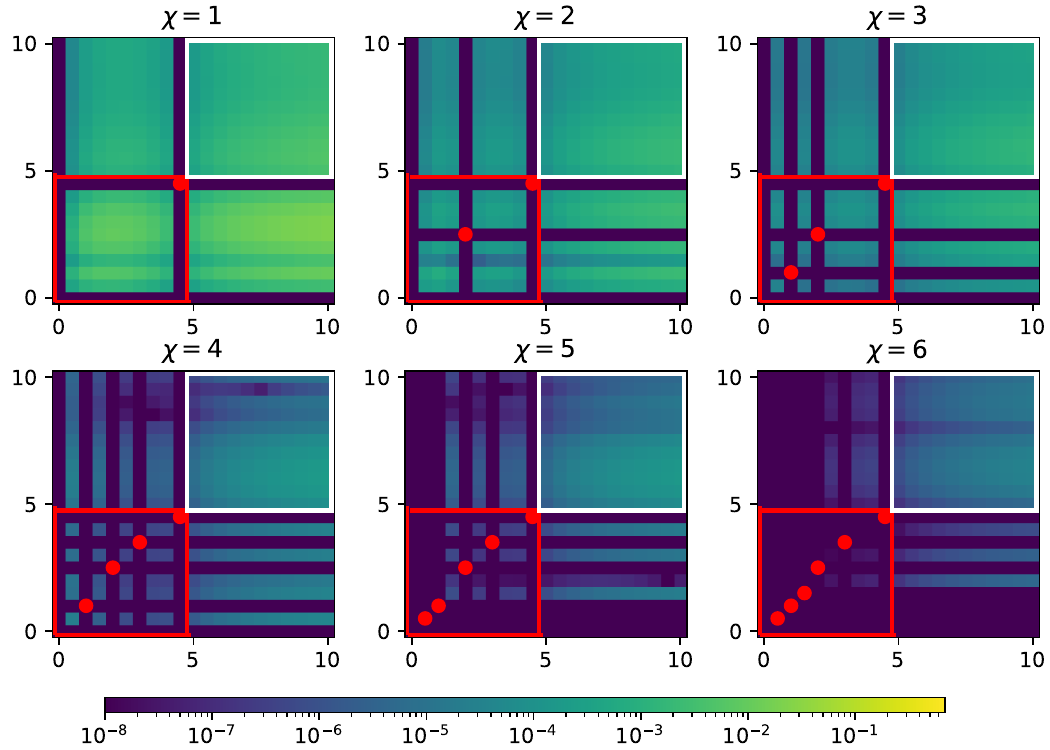}}
	\caption{Error versus $x$ and $y$ at different stages of the $L$-shape Cross-Extrapolation.
	Same function as Fig.\ref{fig:annex_crossi}. The pivots are only added inside the red square and we seek to extrapolate the function in the white square.}
	\label{fig:annex_basic}
\end{figure}

\begin{figure*}
	\centerline{
		\includegraphics[scale=0.5]{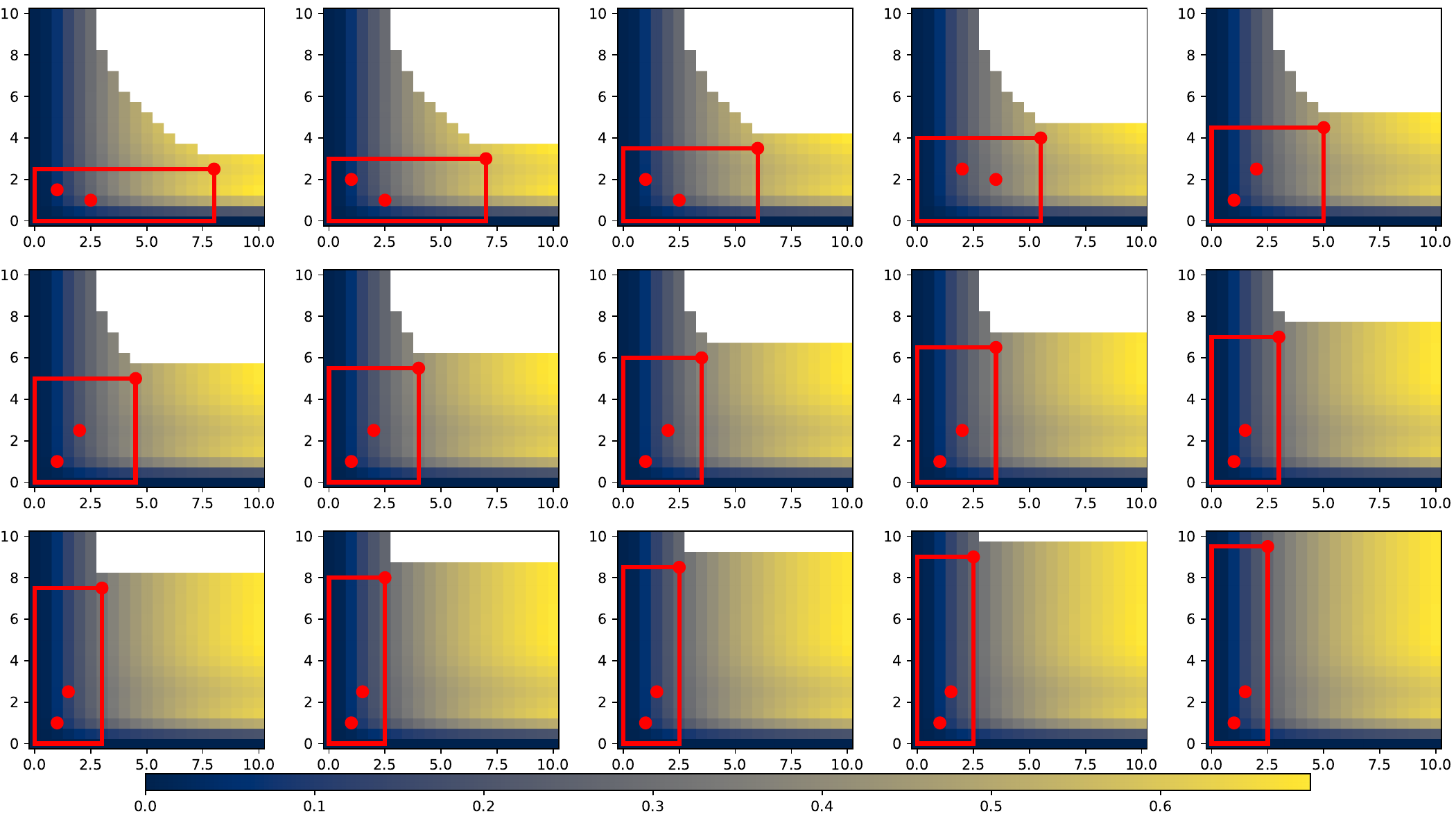}}
	\caption{Extrapolated value of $\ftoy(x,y)$ at different stages of the general-shape Cross-Extrapolation.
		Same function as Fig.\ref{fig:annex_crossi}. The color map correspond to the values of $\ftoy$ known or extrapolated. Values in white (initially for $xy>25$) are unknown and recovered perfoming the extrapolation line by line (the panels progress from left to right and then top to bottom). The pivots (red dots) are selected inside the red rectangle.}
	\label{fig:annex_optimized}
\end{figure*}

In this appendix, we show a step by step illustrations of the different algorithms
in the context of our toy function Eq.\eqref{eq:f}. The figures show the error of the approximation at
different stage of the calculation upon increasing the rank $\chi$. The red circles show the positions of the pivots
and the function has been discretized on a coarse $20\times 20$ grid (for visibility).
\begin{itemize}
	\item Fig.~\ref{fig:annex_crossi} shows the cross-interpolation algorithm (the function is known everywhere,
	${\cal D}_{\rm sub} = {\cal D}$). The pivots are added one by one by selecting the position where the error is maximum. This strategy is known as the Adaptative Cross Approximation (ACA) \cite{ACA2000} and is near-optimal.
	\item Fig.~\ref{fig:annex_basic} shows the $L$-shape cross-extrapolation algorithm for the same function. Here the pivots can only be inside the red square and we seek an extrapolation inside the white square. We observe that the algorithm converges, although not as fast as cross-interpolation.
	\item Fig.~\ref{fig:annex_optimized} shows the general-shape cross-extrapolation algorithm for the same function using the hyperbolic domain $xy < 25$ and $\chi=3$. The different panels (from left to right then top to bottom) show the line by line reconstruction procedure. The function is unknown in the white region of the top left panel.  In order to perform the extrapolation with another values of $\chi$, the full procedure needs to be redone.
\end{itemize}

\section{A remark on left and right currents.}
The calculations shown in the main text correspond to the current $I^l(U,t)$
flowing from the left leads to the quantum dot. However, the current $I^r(U,t)$
flowing from the right lead is also available. In this appendix, we show that the additional information can be useful.

\begin{figure}
	\centerline{
		\includegraphics[scale=0.40]{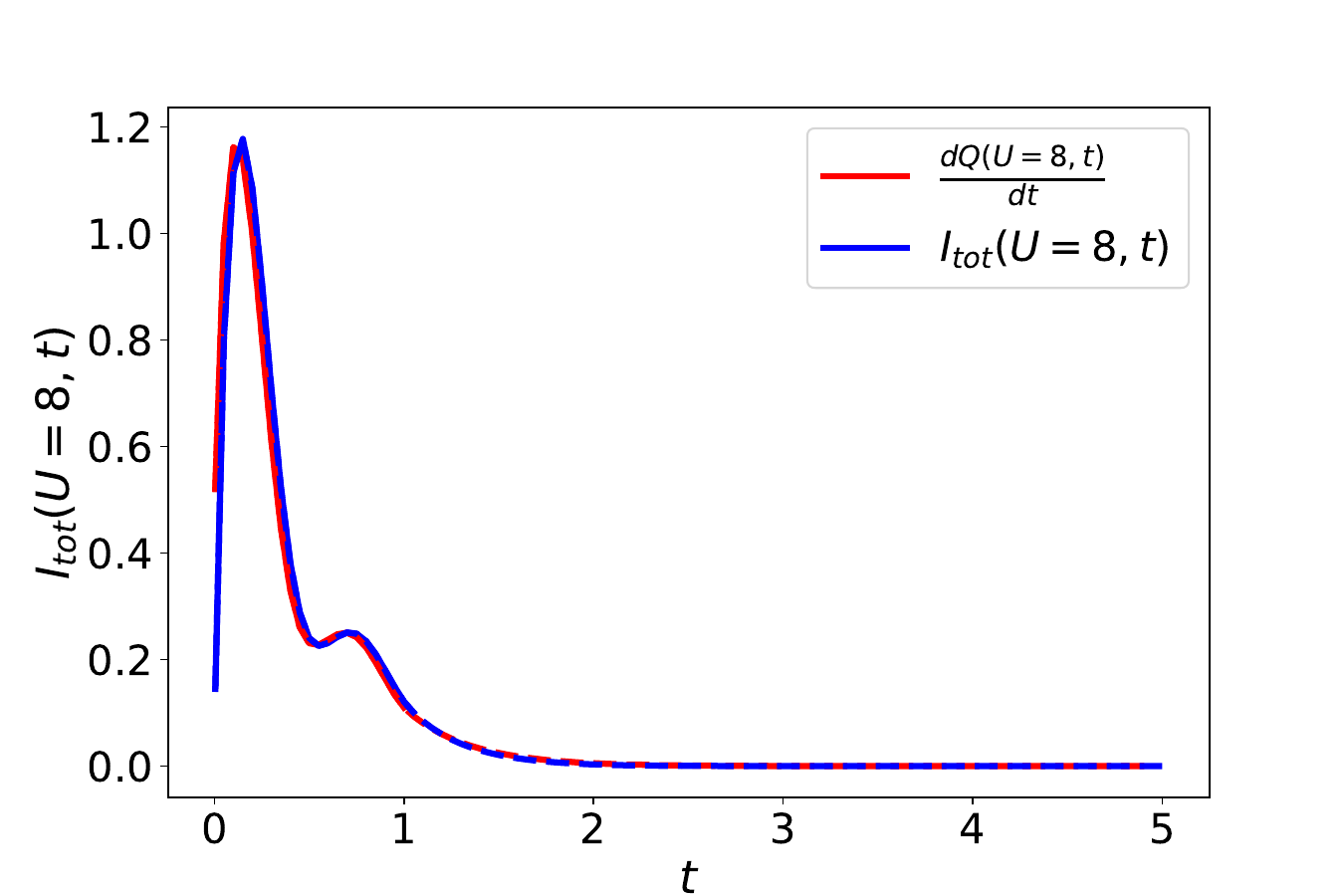}
	}
	\caption{Current conservation for $\alpha=\epsilon_d=V_b=0$, the derivative of the charge (in red) and the total current (sum of left and right current, in blue) at $U=8$ as a function of time. Plane lines correspond to the simple summation and continue into dashed lines in the extrapolation regions. Both extrapolations are in good agreement and compatible with the predicted error bars.}
	\label{fig:annex_conserv}
\end{figure}

First, current conservation provides an independent non-trivial test of the accuracy of the cross-extrapolation. Current conservation reads,
\begin{equation}
\label{eq:cur_conservation}
\frac{dQ(U,t)}{dt} = I^l(U,t) + I^r(U,t) \equiv I_{tot}(U,t)
\end{equation}
and indeed, this equality is true order by order,
\begin{equation}
\forall n, \ \ \frac{dQ_n(U,t)}{dt} = I^l_n(U,t) + I^r_n(U,t)
\end{equation}
so that where the series converges \ref{eq:cur_conservation} is naturally satisfied; however an extrapolation may violate it.  Figure \ref{fig:annex_conserv} calculates both the right and left hand side of Eq.\eqref{eq:cur_conservation}) for 
$\alpha=\epsilon_d=V_b=0$. We find that both extrapolations are in agreement within our calculated error bar, a non-trivial sanity check.

\begin{figure}
	\centerline{
		\includegraphics[scale=0.39]{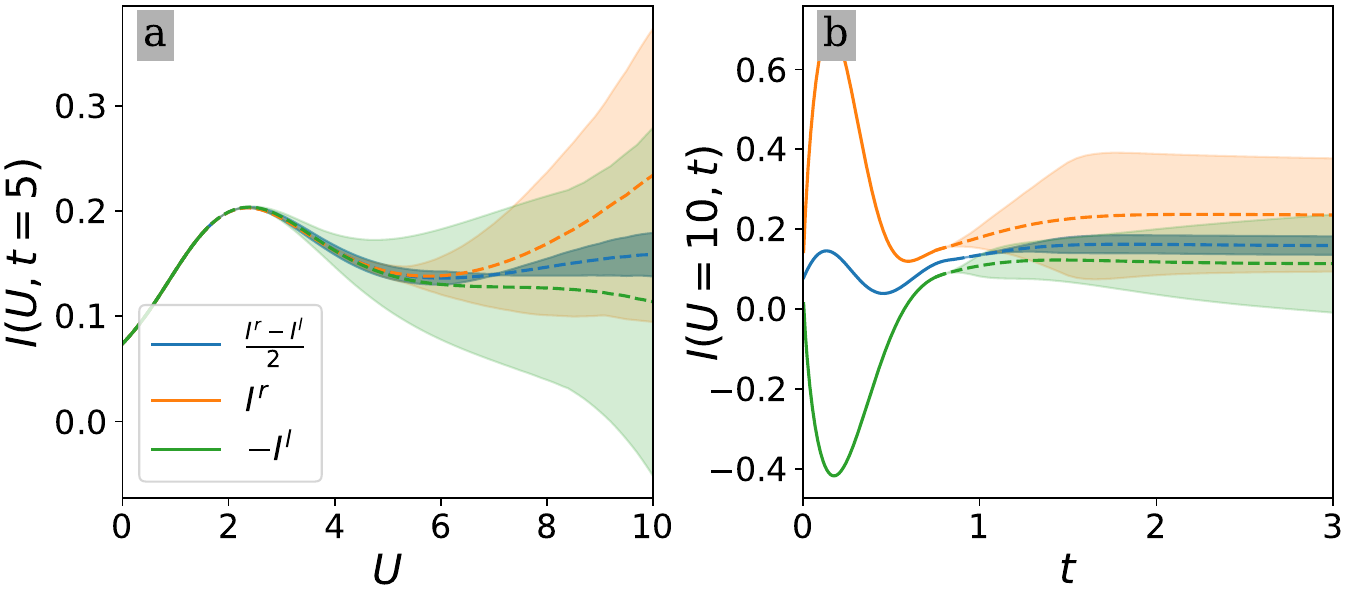}
	}
	\caption{(a) Extrapolation of the difference of the right and left current divided by two (blue), the right current (orange) and the absolute value of the left current (green) using $\epsilon_d=-2$, $\alpha=0$, and $V_b=2$ at $t=5$ as a function of $U$. Error bars are given by the colored area. (b) Same data for $U=10$ as a function of time $t$.}
	\label{fig:annex_cur_diff}
\end{figure}

Second, just after the quench, the system experiments a transient current of electrons flowing  out of the dot in order to reach the new stationary value of the charge $Q$. The currents $I_l$ and $I_r$ are very similar (identical for vanishing bias voltage $V_b=0$). If one is interested in the asymptotic current $I = \lim_{t\rightarrow\infty} I_l(U,t) = -\lim_{t\rightarrow\infty} I_r(U,t)$, then it can be beneficial to extrapolate the difference $I^r-I^l$ that eliminates the short-time transient contribution and facilitate the extrapolation. We explore this idea in Fig.\ref{fig:annex_cur_diff} which shows $I_r$, $-I_l$
and the difference $(I^r-I^l)/2$ versus $U$ ($t=5$, panel a) and $t$ ($U=10$, panel b).
We find that, indeed, the dynamics of $I^r-I^l$ shows less pronounce variations than the other two, yielding significantly smaller error bars in the extrapolation (roughly by a factor $5$ in this example).

\section{image reconstruction}
\begin{figure}
	\centerline{
		\includegraphics[scale=0.22]{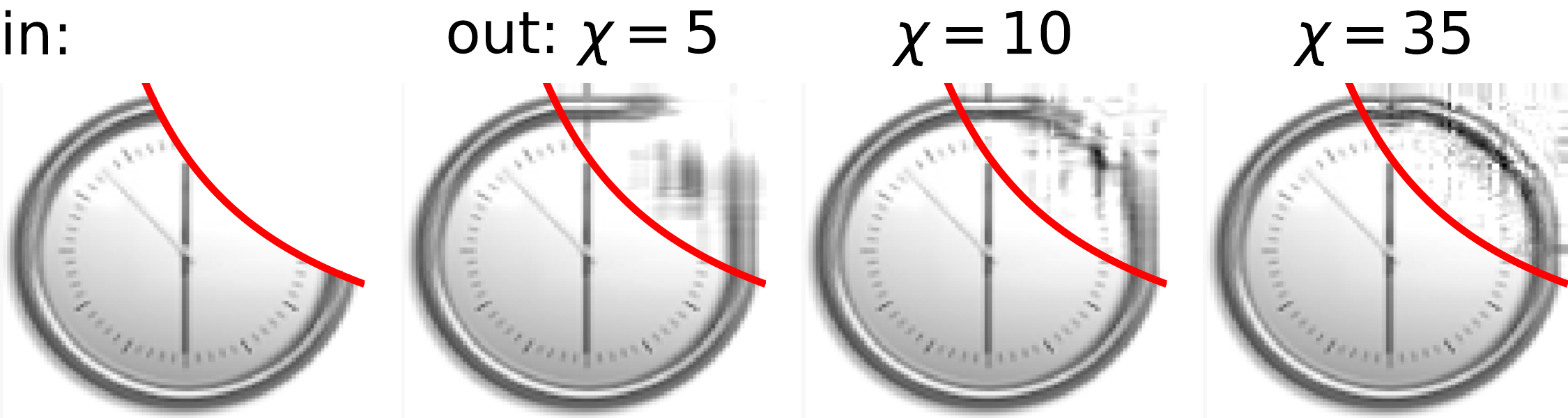} 		
	}
	\centerline{
		\includegraphics[scale=0.35]{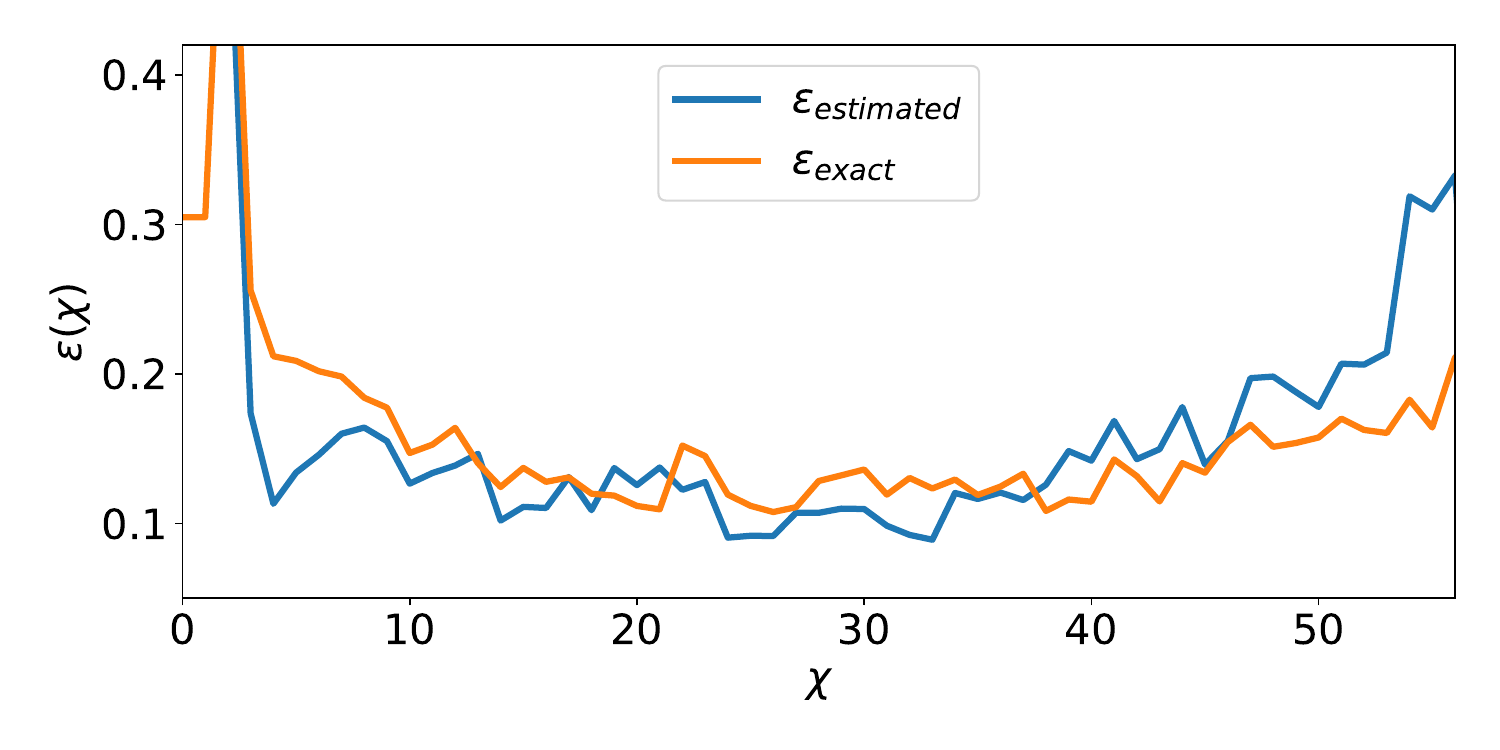}
	}
	\caption{Top left: image of a clock, the part above the red lines has been cropped and needs to be recovered. 
	Top right: image recovered using cross-extrapolation for three different values of $\chi=5,10$ and $35$.  Bottom: estimated norm-2 error per pixel $\epsilon$ (blue) and exact norm-2 error $\epsilon_{\rm exact}$  (orange) between the extrapolated matrix and the original one as a function of the rank. The optimal value of the rank is $\chi\approx30$.}
	\label{fig:image_extrap}
\end{figure}

In this appendix, we explore the possibility to use cross-extrapolation for image reconstruction, i.e. to recover a missing part of  an image provided it exhibits a low-rank structure. We consider a black and white image of a clock which is represented by a 
$100 \times 100$ matrix $M$ with $M_{ij}=0$ (white pixel) and $1$ (black pixel). The image is truncated and we use cross-extrapolation to try and recover the lost region. Figure \ref{fig:image_extrap} shows the extrapolation of the image for different values of the rank $\chi$: while not perfect, the reconstruction works reasonably well for $\chi=35$ with 90\% of the pixels reconstructed correctly.

Importantly, the error of the reconstruction can be monitored as well. 
For an extrapolated matrix $M^\chi$, we use the norm-2 error per pixel
\begin{equation}
\epsilon_{\rm exact} = \frac{1}{\sqrt{S}} |M-M^\chi|_2 = \frac{1}{\sqrt{S}} \left[\sum_{ij} |M_{ij}-M^\chi_{ij}|^2\right]^{1/2}
\end{equation}
to monitor the quality of the reconstruction and $\epsilon = \frac{1}{\sqrt{S}} |M^\chi-M^{\chi+1}|_2$ as our estimate of this error ($S$ is the number of missing pixels).  We find that our estimated error is in good agreement with the exact one.

\end{document}